\pdfoutput=1

\documentclass[pra,aps,superscriptaddress,notitlepage,nofootinbib,floatfix]{revtex4-2}
\usepackage[utf8]{inputenc}
\usepackage{bm}% bold math
\usepackage{bbm}
\usepackage{epsfig}
\usepackage{wrapfig}
\usepackage{float}
\usepackage{mathrsfs}
\usepackage{amsmath,amssymb,amsfonts}
\usepackage{graphicx,xcolor}
\usepackage{comment}

\begin{document}

%%%% Article title to be placed here
\title{Fundamental Physics, Existential Risks and Human Futures}
\author{Adrian Kent}

\address{Centre for Quantum Information and Foundations, DAMTP, University of Cambridge, Wilberforce Road, Cambridge, CB3 0WA, United Kingdom 
\\ and \\
Perimeter Institute for Theoretical Physics, \mbox{31 Caroline Street N}, Waterloo, ON N2L 2Y5, Canada}

%%%% Subject entries to be placed here %%%%
%\subject{Quantum foundations, Gravity, Quantum information, Consciousness, Existential risks, Human futures}

%%%% Keyword entries to be placed here %%%%

%%%% Insert corresponding author and its email address}
%\corres{Adrian Kent\\
\email{apak@cam.ac.uk}

%%%% Abstract text to be placed here %%%%%%%%%%%%
\begin{abstract}
Over the past 25 years, I have been involved in some intriguing developments in the foundations of physics, exploring the quantum reality problem, the relationship between quantum theory and gravity and the interplay between consciousness and physical laws. 
These investigations make it plausible that we will find physics beyond quantum theory, potentially including both new evolution laws and new types of measurement.
There is also a significant chance they could have potentially transformative impact on information processing and on the development of and our future with AI.    
\end{abstract}
\maketitle
%%%%%%%%%%%%%%%%%%%%%%%%%%%

%%%%%%%%%% Insert the texts which can accomdate on firstpage in the tag "fmtext" %%%%%

\section{25 years on}
In my 2000 contribution \cite{kent2000night}, I focused on two central themes: the unresolved quantum measurement problem and the elusive scientific theory of consciousness.
I argued that the quantum measurement problem is a genuine and fundamental
problem in theoretical physics, and against the idea that the once-orthodox Copenhagen
view of quantum theory gives a consistent
and satisfactory metaphysics.   We need either a reformulation of quantum theory
or a deeper theory that gives a unified account of microscopic quantum physics and macroscopic
\mbox{(quasi-)}classical physics.   I pointed towards the dynamical collapse models developed by
Ghirardi, Rimini, Weber, Pearle \cite{ghirardi1986unified,GPR90} 
and others as the then most promising ideas in this direction.  
I also argued we are lacking a fundamental scientific theory of consciousness
that fits with the rest of theoretical physics as we understand it.   The arguments I 
summarized go back to William James \cite{jamesautomata}; relevant reasons to 
be sceptical of reductive materialism go back at least to Democritus (see
\cite{schrodinger2014nature}); David Chalmers \cite{chalmers1996conscious}
gave arguments for a hard problem of consciousness, with which this stance also aligns.  

In summary, we cannot claim a complete
account of physics without including consciousness, since it gives all the evidence for our
physical theories.  We can nonetheless get an (apparently, presently) completely satisfactory
account of material physics by invoking the principle of psycho-physical
parallelism, according to which every conscious perception corresponds to physical events or
processes in (presumably, generally) our brains, which is broadly supported by a wealth
of neuroscientific data.   But psycho-physical parallelism combined with standard material physics implies consciousness is an epiphenomenon.   This makes it hard to understand
why it is there at all, in a universe that would be equally logically consistent without it, and
particularly hard to understand how it has all the properties that make it appear an evolutionarily finely honed survival mechanism, since -- if genuinely epiphenomenal -- it has no independent effect on the material world.

Both arguments invited readers to consider problems
with what I took to be the mainstream views -- some version of Copenhagen quantum theory and
some materialist view of consciousness that denies any fundamental hard problem -- and to 
take alternatives seriously.  This only mildly assertive approach reflected 
genuine adogmatism \footnote{Though also, I didn't have tenure.}: not everything we see as a fundamental scientific problem is necessarily solvable; 
every perspective on consciousness, in particular, is so problematic as to seem incredible.  
Still, my strong hunch was (and still is) that experiment will one day reveal some 
deeper physics underlying quantum theory.  I also think a truly unified and complete description
of nature will include consciousness as a fundamental natural phenomenon somehow -- in 
ways we probably have not begun to conceptualize -- related to mass, electromagnetism,
gravity and the rest.   

We are invited in this volume (I take it) to discuss
how far our visions have been realised, and if, how and why they have changed.  
Space precludes reviewing much interesting and relevant
work by others, so this will necessarily be very much a personal perspective.

\section{What actually is the mainstream view of quantum theory?}
What do physicists really think about quantum theory? Polls taken at quantum foundations conferences offer some clues, though they come with their own biases.
The data in\cite{SCHLOSSHAUER2013222} suggests that the Copenhagen interpretation
was still highly influential in 2013, as was the many-worlds
interpretation. 
My impression is that in some significant physics research communities (for example
string theory and cosmology) many-worlds ideas now dominate.
It should be stressed though that "the Copenhagen interpretation" and "the many-worlds
interpretation" have increasingly become umbrella terms, characterizing broad stances about
fundamental physics rather than precisely formulated theories. 
In particular, there are many, many incompatible ideas (see e.g. \cite{kent2010one,mwbook})
about what many-worlds quantum theory might mean.
From time to time I've pointed out problems with several of the most prominent, 
including co-editing a volume \cite{mwbook} where I succeeded, at least, in persuading
colleagues to frame the title as a question.  

Without rehearsing all the arguments pro and con, let me touch on just one issue raised in \cite{kent2010one} here.
Whatever their other problems, standard "one-world" versions of quantum theory make clear,
scientifically testable statements about sequences of the apparently random outcomes
of quantum experiments.   We can
understand these in terms of Kolmogorov's notion of algorithmic compressibility.  
In this language, the hypothesis that a two-outcome experiment has a $1/3$ probability of giving outcome $1$
means that if you repeat the experiment many times then (a) you expect to be able to compress the list of $N$ outcomes to a list of length $N H$, where $H = - 1/3 \log (1/3) - 2/3 \log (2/3)< 1$ is the Shannon entropy of the outcome probability distribution, (b) you do not expect to be able to compress it significantly
further.   If either prediction fails, then you should lose credence in the hypothesis. 
Many-worlds quantum theory gives a different hypothesis: all possible sequences of outcomes will
really occur and be observed by a future successor of yours in some world.
All these successors are equally real: none of them has a stronger claim to be "you".  
If "you" repeat the experiment many times, you are generating many successors observing
different sequences, each of whom is unaware of the others.   Whatever sequence any of
them observes is consistent with the theory's prediction about reality, which is thus unfalsifiable. 

As my earlier article noted, the Copenhagen interpretation precludes a unified scientific theory 
encompassing microscopic and macroscopic physics, let alone cosmology.
Many-worlds quantum theory proposes a unification -- unitary quantum mechanics describes both microscopic and macroscopic, and so macroscopic superpositions persist.  
But its basic premise precludes confirmable or refutable
scientific theories, and this seems impossible to avoid given the basic hypothesis.\footnote{Though for some attempts see \cite{mwbook}.} 

The universe is a strange place ; perhaps no unified theory can fully describe it.
If there is one, it might not be scientifically confirmable by observers within it.
But we have made great progress in understanding it by seeking theories
that are progressively more unified {\it and} scientifically confirmable.
Methodologically, it makes sense to
continue with standard scientific assumptions that have proven so fruitful, unless
and until we have compelling reasons to think we have reached the limits of their validity.
My guess is also that this will succeed: I would find a universe that completely resists scientific
investigation, or one that can only be described by a patchwork of partial and incompatible
theories, less surprising than one that holds out so much hope for science
and unification, yet ultimately dashes these hopes.    

\section{Three paths to physics beyond quantum theory}

\subsection{Beables and new dynamics}

{\bf \color{blue} 1. Beables}  \qquad  As already noted, the Copenhagen interpretation gives no precise way
of unifying microscopic and macroscopic physics, which we need.    
Its implications were articulated most precisely by Bell \cite{bell1976theory,bell1987beables}. 
We need a mathematical formalism characterizing what, exactly, quantum probabilities
are probabilities of.   As Bell put it, quantum theory is presented as a theory of
observables, without telling us who or what qualifies as an observer.
We need, he argued, a theory of {\it be}-ables \cite{bell1976theory}, mathematically well-defined
entities that characterize physical reality, within which observers, 
macroscopic objects, galaxies and so on
are defined.  On this view, none of the latter plays a fundamental role
or necessarily has a precise definition, but the beables do and must.   
Mathematically, they define the sample space for quantum probabilities. 

Bell pointed to de Broglie-Bohm theory \cite{debroglie,bohm} and dynamical collapse models \cite{ghirardi1986unified} as interesting
examples of beable extensions of quantum theory.   Much theoretical effort has gone
into developing dynamical collapse models in the last 25 years, but there remains no
satisfactory relativistic field-theoretic version.  Even more effort has gone into devising and
implementing experiments to test the most studied models (Ghirardi-Rimini-Weber-Pearle (GRWP)
mass-dependent spontaneous localization \cite{GPR90}), narrowing the parameter
window \cite{carlesso2022present} 
in which they remain viable to the point where (even if we set aside most
people's initial low credence) it would begin to seem a little conspiratorial if
nature had chosen model parameters in the relatively small ranges that technology
does not yet allow us to test.   So I tend now to see GRWP models to
date as interesting and well-motivated modifications of quantum theory that are not only
very likely incorrect, but probably will not even turn out to be going in the right direction,
yet should encourage us to explore further.   
De Broglie-Bohm theory, similarly, has no satisfactory relativistic field-theoretic version
and, in its usual formulation, is not experimentally distinguishable from Copenhagen quantum
theory.\footnote{Though it can be made so.\cite{Valentini1992}}
It also feels, in comparison to GRWP models, clunkily mathematically hybrid, combining the mathematical formalism of quantum theory with classical trajectories of point-like particles in a way that lacks the elegance we have come to expect from successful new physical theories.  

So we need better beable theories.  One idea I've been developing   \cite{kent2015lorentzian,kent2017quantum} combines insights from
several approaches to quantum theory, including collapse models, the physics of decoherence, and ideas in which initial and final states both play a fundamental role.   The essential idea is that, in (fairly standard) cosmological models in which the universe expands forever, once we have a theory of the initial conditions and the unitary dynamics, a complete beable description of reality could be relatively simply and elegantly reconstructed from the asymptotic late time state of the quantized electromagnetic or gravitational field.  This offers the hope of an explicitly relativistic model that (unlike de Broglie-Bohm theory) fits naturally with the framework of relativistic quantum theory.   Of course, even if such a model is fully developed, it may not be correct.  But it illustrates at least there is significant scope for new beable theories.
\newline

{\bf \color{blue} 2. New dynamics from beables} \qquad  Even an elegant beable 
model built thus on standard unitary quantum theory has a counter-intuitively unaesthetic 
feature: the beables are the fundamental building blocks of reality, yet inert.   
The mathematics of quantum theory defines the space of beables {\it and} the
probability distribution of beable configurations.  But it need not. 
We can define consistent theories \cite{kent2013beable} in which the probability distribution of
a beable configuration depends on intrinsic properties of that configuration
as well as on the quantum dynamics, even if we continue to restrict ourselves to standard
quantum theory, in which those dynamics are determined by the initial state and
a Hamiltonian that determines its evolution.   Most theories defined in this 
way will look rather ad hoc -- but the universe only needs one (which
might appear elegant only after reframing, or, more likely, which might
be only an approximation to an elegant underlying theory).   
The problem, if this line of thought is fruitful, is finding it, starting as we 
do from standard quantum theory.    

Happily, framing the problem points not only to a solution, but also to a 
vital scientific project that is motivated whether or not one takes beables seriously. 
A "beable-guided" theory \cite{kent2013beable} that modifies standard quantum theory tells us that the 
probabilities of some sequences of events are different from those we obtain
from the initial conditions and the quantum evolution laws.   This is not because
we have the wrong initial conditions or evolution laws, but because (in such
a theory) initial
conditions and standard evolution laws don't suffice as a description of the universe.
One way of picturing this \cite{kent1998beyond,kent2022hodology} is that the apparently random dice determining quantum events
are actually collectively (not individually) biased to steer the universe towards 
some evolution paths and away from others.   We do not know where to look for evidence 
of these "hodological" (from {\it hodos}, Greek for path) models
at the level of fundamental beables, absent a clear intuition about the form 
of a fundamental "beable-guided" theory. 
But we can, and should, look for evidence that large-scale phenomena -- in particular,
cosmological events -- are better described by non-standard hodological modifications of 
quantum theory than by standard quantum theory.   

A common initial concern is that this takes us beyond the realm of science. 
In principle, one could devise a beable-guided theory that steers the universe towards any
evolutionary path; in particular, it could steer the universe towards the very specific
path we have observed, including the distribution of stars and galaxies and even the 
specific evolutionary history of life on Earth.    Suggesting that things turned out
the way they have because one has a theory that says (post hoc) in precise detail
that they must have certainly would not contribute to science!

As in contrasting one-world and many-worlds quantum theory, 
algorithmic compressibility, specifically its development 
by Solomonoff \cite{solomonoff1964formal} into an algorithmic formulation of scientific induction, show
us how to separate scientifically empty models from fruitful ones, and how
to work towards confirming or refuting the latter.   
Formally, the key idea \cite{kent2022hodology} here is to use
the principle of minimum description length (MDL) for hypothesis 
identification \cite{rissanen1978modeling}, according to which the best hypothesis to fit the data is the one that minimizes the sum of the length of the program required to frame the hypothesis and the length of the string required to characterize the data given the hypothesis.
Informally, cosmological models that modify quantum evolution relatively elegantly,
using simple additional rules with few parameters, deserve some credence, 
and can and should be tested against the standard paradigm.    
Models that spell out in fine detail how the universe evolved are given 
essentially no credence, because specifying them takes a very large amount of information
(in the extreme case, recapitulating the actual history of the universe).
There are only a finite number of models of any given description length, and 
we can confirm or refute these with a finite amount of observational data.  

Recent hints \cite{desicollaboration2025extendeddarkenergyanalysis} that dark energy may vary over time and space -- a feature that 
could easily be incorporated into a hodological model, which might or might not 
require few parameters to fit the data well -- add significantly to the 
motivation for this program.  
Granted, cosmological data is hard to interpret, and cosmologists continue to 
argue about the strength of evidence for or against standard cosmological 
theories.  Also, Solomonoff induction is defined with respect to a given
computing model and language, which leaves some (albeit finite) uncertainty about how to 
apply it in practice to science framed in a mixture of human language and mathematics.   
These issues make implementing this program somewhat more complex, but do not alter
the key point: we can certainly produce simple hodological cosmological models that do not fit
into the Newtonian initial-causes-suffice paradigm, and we should be trying to 
test that paradigm against them.    
Even if one thinks of these models only as foils rather than serious contenders, 
they allow us to parametrize and make precise the extent to which we have confirmed
the standard paradigm.   

An illustration of the effects of a simple hodological rule in a simple toy model is given in 
Fig.~1.  As discussed there, if we saw such effects our credence in a hodological rule
similar to the one given would be greatly increased, and our credence in a standard Newtonian
rule correspondingly diminished.

\subsection{Gravity}

The second reason for looking beyond quantum theory is that we still don't have a conceptually satisfactory quantum theory of gravity, nor any evidence that gravity is quantized.  
Many physicists have come to believe we don't need evidence because 
there is no consistent way of combining a 
classical theory like general relativity with quantum theory.
The argument here is that any coupling
between classical and quantum degrees of freedom would allow us (at least in 
principle) to obtain "too much information" about quantum states, implying
faster-than-light signalling.   However, arguments for
this \cite{eppley1977necessity} have been refuted \cite{kent2018simple}.  In fact, it turns out there
are very many logically consistent ways of extending quantum theory that allow 
non-quantum measurements \cite{kent2005nonlinearity,Kent2023,fedida2024mixture} that give more information
about quantum systems than standard quantum measurement theory permits.

More recently, it has been suggested \cite{Masanes2023} that such measurements are inconsistent with the
other postulates of quantum theory.  
This argument relies not on faster-than-light signalling but
on the related but distinct so-called quantum no-signalling principle, 
which says that operations on one isolated subsystem should have no measurable 
effect (at any time) on another.   
This is sometimes expressed as the principle that physical influences must
have physical carriers.     
In the context of general relativity, though, there are no truly isolated 
subsystems: matter anywhere influences matter everywhere through its 
gravitational field, i.e., its action on space-time, which indeed serves
as a physical carrier for physical influences. 
It's certainly possible to imagine that, in some theory unifying quantum theory
and general relativity, additional (presently unspecified) degrees of freedom
carry additional information alongside, and presumably intertwined with, the
information that space-time carries in special relativity, and 
that non-quantum measurements use this additional information.  
For this reason, I place higher credence than most of my colleagues 
in the possibility that future physics will go beyond standard quantum measurement theory \footnote{Renato Renner and I have bet on whether this will happen by 2040, staking
100 cases of wine (Renato) against one (me).} -- which would potentially have a dramatic
effect on quantum computing and quantum cryptography as well as fundamental physics (see below).  

These arguments have become much more pertinent because of one of the most exciting recent developments in fundamental physics, the realization that the quantum nature of gravity can be tested in the low-energy regime by finding evidence for or against gravitationally-induced
entanglement or other effects that are predicted by all standard quantum gravity theories 
but not by theories in which the gravitational field is classical.     
The key idea was first discussed by Bose et al. \cite{bose2017spin}
and independently by Marletto and Vedral \cite{marletto2017gravitationally}.   
Although the relevant experiments are not feasible with present technology, they are
much closer to feasibility than previous proposals relying on Planck energy effects.  
They have added impetus to work (e.g. \cite{Oppenheim2023,tilloy2016sourcing}) 
on specific hybrid theories in which
a classical gravitational field is coupled to quantum matter, as well as to the general 
possibilities just mentioned.   
Debate continues (e.g. \cite{huggett2023quantum,rydving2021gedanken,martin2023gravity}) 
over exactly what these experiments or variants \cite{Kent2021testingthe} would test.   
The initial hope that they might give a clean and definitive test of whether 
gravity is quantum, with essentially no additional assumptions, seems over-optimistic.
However, they would either refute or confirm the most interesting 
alternative types of model identified so far, and hence either refute or
give strong new evidence for quantum gravity.
Moreover, simpler versions of the experiments, which are likely to be 
feasible sooner, will already refute or confirm some interesting 
alternatives \cite{kent2021testing,kent2021quantum}.    

\subsection{Consciousness}

Writing about consciousness and physics 25 years ago felt risky, and it felt positively foolhardy to promote William James' argument \cite{jamesautomata} against consciousness
being an epiphenomenon, a byproduct with no causal influence.
As James pointed out, if so, it is implausibly fine-tuned, appearing 
very well developed to reinforce
evolutionary advantageous behaviours while actually producing a mere narrative, laden with attractions and aversions that appear to motivate but
have no independent effect on the material organism.    
It has been heartening to see an explosion of interest over the last 
couple of decades in consciousness as a fundamental physics problem, 
guided, inter alia, by David Chalmers' beautifully lucid expositions \cite{chalmers1996conscious}
of the motivations for and difficulties with every line of thought, 
stimulated further by attempts at an "integrated information 
theory" \cite{tononi2012phi,tononischolarpedia} of consciousness, including a revival of interest (e.g. \cite{chalmers2015panpsychism,goff2019galileo}) in 
the physical arguments for panpsychism promoted by Russell \cite{russell1921analysis}, Eddington \cite{eddington1928nature}
and others, and also including a serious proposal to test the 
implications of James' argument by looking \cite{neven2021robots} for deviations from 
quantum theory on a quantum computer.   

Let me summarize where I think we currently stand.  Some argue that 
that creatures that have evolved to interpret and reason 
about the world and their own interactions with it necessarily must be conscious,
and that we, with our brains, necessarily must be conscious in the way we are.
They deny the logical possibility of philosophical zombies, inhabiting an
alternative universe with a material world identical to ours (starting from
the same state, following the same laws, with the same random quantum
events) but without any consciousness.   To many others, including me,
this is incoherent: it is an objective fact that I (and I'm sure you, 
and almost all animals) have subjective experiences, but not a fact
that follows from quantum theory or general relativity.   It's logically
consistent, given what we understand about physics and consciousness
at present, to imagine a universe with the same material evolution 
in which those experiences were different,
or absent. 

Dialogue between these camps can be difficult.  Many people
seem to join one early in their intellectual life and find their 
conceptual framework allows them to translate what the other camp seems
to be saying as some form of obvious error or failure to appreciate
elementary points.  Only a minority come to appreciate that
their own position does actually, like every stance on consciousness,
have difficulties.   For example, accepting James' argument against
epiphenomenal consciousness, I am still struggling \cite{kent2016quanta,kent2020toy,kent2021beyond} even to sketch the possible form of a non-epiphenomenal
model that could satisfactorily explain why our consciousness appears 
evolutionarily well-adapted.

When thoughtful people disagree so radically on a fundamental point, on 
which every articulated position appears problematic on careful analysis, one ought rationally
to broaden one's credence distribution.   James' argument, arguments for
panpsychism or panprotopsychism, and arguments for dualism all suggest there should be new
dynamical laws associated with consciousness.  Arguments for epiphenomenalism or illusionism
suggest there should not.  So, arguably, does the fact that current material physics 
does indeed describe the dynamics of living creatures very well: we have seen no dramatic
anomalies, so any new dynamical effects would have to be very subtle.
Cashing all this out leaves me with significant, if not overwhelmingly high, credence in new
laws to be found.   

\section{Synthesis}
Three distinct threads -- quantum reality, gravity, and consciousness -- give completely different motivations for exploring physics beyond quantum theory.
We also have a panoply of examples of ways in which quantum theory can consistently be
modified.   These threads could weave together converge into a single tapestry, revealing a universe far richer and more interconnected than most physicists 
have so far imagined.   For example, the unification of quantum
theory and gravity could involve a hodological model whose predictions are most 
easily tested by cosmological observation.    Or, as Penrose has suggested \cite{penrose}, resolving
the quantum measurement problem might involve collapse mechanisms associated with
both gravity and consciousness.    Or even, much much more speculatively, 
fleshing out arguments by Nagel \cite{nagel2012mind},
the emergence of consciousness in the universe might need a hodological explanation.
But the arguments need not connect and of course they need not all be right (even if one is). 

We can and should test separately the ideas motivated by each line of thought. 
Much effort is now going into developing the technology needed to test quantum
gravity against alternatives, and into theoretical ideas that might allow 
easier or different tests.  
Proposals that would test in some way the relationship between consciousness
and quantum theory include: the aforementioned searches for anomalous outputs
from quantum computers; work on collapse models that implement Wigner's
hypothesis that consciousness collapses wave functions, using explicit
mathematical models for measures of consciousness; experiments involving
living organisms; Bell experiments with well-separated human observers
directly observing the outcomes in each wing.
I hope to build a collaboration testing a good range of hodological cosmological models. 

We should take on board the broader moral, though: new physics underlying quantum theory
could manifest itself in many ways, most of which we likely have not yet envisaged.
This motivates a much broader
and more systematic "stress-testing" of quantum
theory in untested regimes, and an active search for dynamical anomalies.  
For me, this is the most important project in science.
We could finally go beyond the Newtonian paradigm; we could find
a deeper theory underlying quantum theory and relativity; we could
gain new fundamental insight into the nature of consciousness and
our relation to the universe.   And we could also possibly transform
information technology and the evolution of intelligence on (and beyond)
Earth, as the next section discusses.   

\subsection{Fundamental physics and human futures}

My earlier essay closed with the comment that the Editor of the corresponding volume in 2999
would very likely be able to solicit contributions from extraterrestrial and/or AI colleagues.   Today, that future feels much closer.
Like most, I didn't foresee that AI contributions would already be possible
(if arguably not yet quite as interesting as human ones) in 2025. 
It now looks very plausible that there will be general-purpose
superhuman artificial intelligence in the coming decades.
Not everyone is persuaded (e.g. \cite{blackwell2024moral}).
On the other hand, many experts think it will be much sooner \cite{grace2024thousands};
indeed, serious efforts seem under way to promote 
a Manhattan-style project to achieve it in the next few years.
Let us accept that the hypothesis of artificial general-purpose intelligence in a few decades
deserves significant credence: it does not
ultimately matter so much for the argument whether the credence is $1\%$ or $99\%$.  
I am now going to set out arguments, with the caveat that
most need much further and more thoughtful
analysis.   This is a manifesto for a research programme,
a key part of which is to scrutinize the plausibility of each hypothesis.

We start with the hypothesis that
humans have perhaps 30 years (and maybe significantly fewer) left to steer fundamental physics research, before AI takes over.   The only possible discoveries that could
have a longer-term impact on our well-being are those that affect
the development of AI itself.   Extending the theory of elementary particles
would be scientifically fascinating, of course, but seems very unlikely to
have any technological impact in 30 years.   Nuclear physics did, of course, have 
enormous impact.    A transformatively new way of generating 
energy could have some impact, for better or worse -- for example
the cold fusion dreamed of by mavericks might, hypothetically, allow 
effectively unlimited cheap clean energy (and so cheaper and less environmentally
damaging AI data centres) but also cheap and potentially
devastating new radiation weapons.   But (pace \cite{Metzler_2024}) there seems no 
plausible reason to expect such a development. 
The progress of sub-atomic
physics since the 1930s and essentially all serious theoretical ideas about particle physics beyond the standard model suggest we shouldn't expect new ways of liberating energy from any
other development in this area either.   

The other possibility that could profoundly impact the development of AI and other
key technologies is a 
transformative development in information processing.
This {\it is} a more plausible
consequence of the possible new physics beyond quantum theory we've discussed. 
Quantum computing is believed to be more powerful than classical computing
at some significant tasks, including factorisation, simulation of quantum 
systems and (though with only a square root speed-up) searching.   
Many generalisations of quantum theory would, in principle, give models
of computation that are significantly more powerful than quantum theory.
For example, in models of computing 
using a perfect implementation of a nonlinear version of quantum 
theory we would have\cite{abrams1998nonlinear} P=NP, implying oracle-like computing power, with which obtaining
the solution to what (in our current understanding is) an intractably
hard problem would not be significantly harder than verifying a given solution \cite{aaronson2016p}. 
It seems unlikely any post-quantum physics would give us the effectively infinite precision
control that these models require, but not so implausible that it could give us a significantly
stronger model of computing, and perhaps thus a significantly more powerful
form of machine learning, than quantum theory does.  
It could similarly transform our understanding of more general physical learning machines \cite{Milburn_2022}.   

We thus have a (perhaps small, though I suspect significantly larger than most physicists consider)
probability $p$ of a future fundamental physics discovery that would have a 
very large impact $I$ on human welfare and on the future development of intelligence.
To the extent that we think humans will be able to beneficially steer the development of AI 
(on which, to be clear, expert opinion ranges from hubris to despair) we should much 
prefer that this discovery is made by humans, in the phase where we are still 
designing AIs and integrating value alignment and guardrails, rather than by AI, in 
the phase where it is self-designing and any attempts at human steering have less
if any effect.   

One of the most significant developments of the last 25 years has been the academic
mainstreaming of research and policy focussed on this type of (maybe) low
probability, (certainly) high impact existential risk and opportunity.
In 2000, when I pointed out major flaws in risk assessments for the speculative
hypothesis \cite{dar1999will,jaffe2000review} that so-called "killer strangelets" might be created in collider
experiments and cause a catastrophic runaway reaction, 
I encountered much resistance from colleagues even to the elementary point \cite{kent2004critical}
that one needs to evaluate risks in proportion to their impact as well
as their probability, meaning that a 
probability bound of $2 \times 10^{-5} $
on accidentally destroying the Earth gives a decidedly
unreassuring bound of $10^5$ in expected present human lives lost.
A truer measure \cite{kent2004critical} of the hypothetical catastrophe allows for future
human lives lost as well as present, giving a bound (obviously with much
more uncertainty) of something very roughly in the region of $\approx 10^{13}$
human lives.   

Thanks to the work of the Cambridge Centre for the Study of 
Existential Risk, Oxford's Future of Humanity Institute, and partner
organizations around the world, the naive arguments offered by CERN scientists \cite{dar1999will} to the effect 
that "$p$ is 
proven small, or at worst comparable to other natural extinction risks, so the experiment
is safe enough" would encounter much
more formidable resistance today.  
This is not to say we would or should now be paralysed by each and
every new risk hypothesis.  The existential risk community have also popularized
and addressed the problem of "Pascal's mugging", a term first coined by
Elizier Yudowsky, in which a plethora of minuscule probability ultra-high-risk and/or -reward
hypotheses threaten rational decision-making.   
Before being too swayed by purported bounds for expected lives or related quantities, we need, at least, to be confident in evaluating a lower bound for $p$, an upper bound for $I$, 
and, crucially, a horizon scan of other hypotheses with similar or higher estimates of $pI$.   
In the case at hand, this means asking, inter alia: What should our credence in alternatives to 
quantum theory be?  That, if there is a theory underlying quantum
theory, it actually gives major advantages for computing and machine learning in principle?
That it gives advantages that could realistically be exploited technologically in the next 30 years? 
That such a theory will be found in the next 50 years?   
That a more human-friendly form of future intelligence is significantly more 
likely if the discovery is made and exploited by humans?   
That careful answers to all of this leave the hypothesis high on the list of concerns worth
addressing, given the several known significant existential risks 
(see e.g. \cite{ord2020precipice}) and many other hypothetical ones?  
While I lack definitive answers, the urgency and plausibility of these questions compel me to advocate for a larger-scale, collaborative effort to explore them.

Provocatively, but seriously, I have set out in Table $1$ a list of the relevant hypotheses, my
considered credences at present, and their impacts if true.
The credences may seem high to many readers.   One reason for that is that I have quite high credence
that the quantum measurement problem requires new physics, low credence that any version of
many-worlds quantum theory can resolve it, and also quite high credence in stances that suggest
new physics associated with consciousness.    If your credences in each of these are enormously
different, it's worth considering that many thoughtful physicists differ, and asking how confident
you should be that your reasoning and intuition are superior to theirs.   (The reverse is true
too, of course: this is why my credences don't approach certainty.)   Another reason is the
general point that historically, scientists have always tended to be overconfident in the 
current paradigm.    
Of course, even taking these points into account, credences can still reasonably differ substantially. 
Note though that even credences $10^{-3}$ smaller than mine would still produce very large values of $pI$
and justify supporting research programmes, if Pascal's mugging considerations turn out 
not to mandate downscaling the effective impact very substantially.\footnote{For comparison, Toby Ord
gave \cite{ord2020precipice} estimates of $10^{-2}$ for the risk of existential catastrophe arising from nuclear war or catastrophic climate change over the next century.   These smallish estimates (now somewhat
upgraded and downgraded, respectively \cite{ordtalk}) certainly weren't meant as arguments against
devoting large resources to mitigating these risks.} The next decades may well determine whether humanity remains the primary driver of scientific discovery -- or cedes that role to the very intelligences we create.  The programmes I have outlined at least increase the chances that we retain more control, for longer, and propel science along a path better
aligned with human values.   

\acknowledgements{I would like to thank my research students and many other collaborators
and colleagues working on the foundations of physics for continually probing the
boundaries of our understanding and making it all so enjoyable and worthwhile; 
the Centre for Quantum Information and Foundations and Perimeter
Institute, my intellectual homes, for fostering so much groundbreaking
foundational research; the Royal Society,
DAMTP, Perimeter Institute and the Foundational Questions
Institute for invaluable support; 
the Cambridge Centre for the Study of Existential Risk
for their pioneering work and continual reminders to focus
on the bigger picture; Wolfson College and Darwin College, 
my welcoming, lively and supportive scholarly communities; Seth Lloyd
for helpful comments. \newline
I acknowledge support from UK-Canada Quantum for Science research collaboration grant OPP640.  This work was supported in part by Perimeter Institute for Theoretical Physics. Research at Perimeter Institute is supported by the Government of Canada through the Department of Innovation, Science and Economic Development and by the Province of Ontario through the Ministry of Research, Innovation and Science.}

\section{Figures \& Tables}

\begin{figure}[!h]
\centering\includegraphics[width=2.5in]{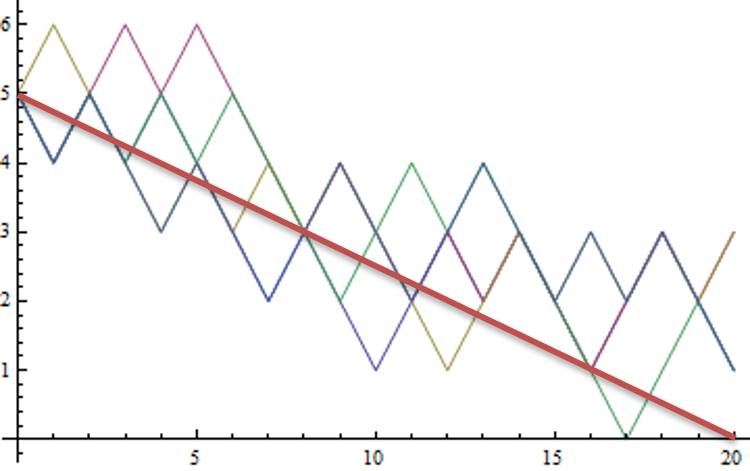}
%%% where xxxxxx name represents "figurename.eps"
\caption{Effect of a hodological path-guided law in a simple Ehrenfest urn model (from \cite{kent2022hodology}).
Ten numbered balls are initially divided equally between two urns.   At each of $20$ steps, 
a random ball is chosen and moved to the opposite urn.  The standard model has no other rule,
and produces fluctuations around equidistribution.   The rule illustrated here weights the 
probability of any sequence depending on its mean square separation from the line 
$n=5-(t/4)$ (red line), producing typical sequences shown by the other coloured paths.
The chance of any such path arising by chance in the original model is so small that, if
it were observed, we would inductively infer a rule similar to the one given. This simple model illustrates how path-guided laws could steer the universe to favour
some paths, challenging the independent random time-step evolution
predicted by standard quantum theory.}
\label{fig_sim}
\end{figure}
\vspace*{-5pt}
\begin{table}[!h]
\caption{Hypotheses, credences, impacts}%%%Table caption goes here
\label{table_example}
\begin{tabular}{|l|l|l|}
%%%The number of columns has to be defined here
\hline
{\bf Hypothesis} & {\bf Credence} & {\bf Impact} \\
\hline
Path-guided evolution laws & 0.25 & Revolution in understanding \\
 & & quantum theory \& cosmology\\
 \hline
Non-quantum gravity & 0.25 & Revolution in understanding \\
&& fundamental physics\\
\hline
"Post-quantum" measurements (PQM) &0.1 & Revolution in understanding \\
&& quantum theory \\
\hline
PQM give much more powerful computing: && \\
in theory& 0.04 & Potential technological revolution \\
 in medium term& 0.01 & Technological and AI revolution \\
 \hline
New evolution laws &0.2  & Revolution in understanding\\
connected with consciousness (EC) & & mind-matter relation \\

\hline
EC allow an effective consciousness meter &0.15 & Policy and ethics revolution \\
\hline
EC give much more powerful computing: && \\
in theory&0.12&Potential technological \& AI revolution \\
in medium term&0.01&Technological \& AI revolution \\
\hline
\end{tabular}
\vspace*{-4pt}
\end{table}%%%End of the table

%%%%%%%%%% Insert bibliography here %%%%%%%%%%%%%%
\newpage
\bibliographystyle{unsrt}
\bibliography{deduped}
%\bibliography{sample,RSTA/postdocbiblioshort,RSTA/library,RSTA/braincollapse,RSTA/collapselocexpt,RSTA/qualia,RSTA/qualia2,RSTA/qualia3,RSTA/photonreality,RSTA/reality}

\end{document}